# Fluorescent oxide nanoparticles adapted to active tips for near-field optics

A Cuche[1], B Masenelli[2,§], G Ledoux[3], D Amans[3], C Dujardin[3], Y Sonnefraud[1,4], P Mélinon[2] and S Huant[1,§]

[1] Institut Néel, CNRS and Université Joseph Fourier, BP 166, 38042 Grenoble, France
[2] Université de Lyon, F-69622, Lyon, France; Université Lyon 1, Villeurbanne; CNRS, UMR5586, Laboratoire de Physique de la Matière Condensée et Nanostructures
[3] Université de Lyon, F-69622, Lyon, France; Université Lyon 1, Villeurbanne; CNRS, UMR5620, Laboratoire de Physico-Chimie des Matériaux Luminescents
[4] Experimental Solid State Physics, Blackett Laboratory, Imperial College London, Prince Consort Road, London SW7 2BZ, United Kingdom

[§] Authors to whom correspondence should be addressed
E-mail : bruno.masenelli@lpmcn.univ-lyon1.fr
serge.huant@grenoble.cnrs.fr

**Abstract.** We present a new kind of fluorescent oxide nanoparticles (NPs) with properties well suited to active-tip based near-field optics. These particles with an average diameter in the 5 to 10 nm range are produced by Low Energy Cluster Beam Deposition (LECBD) from a YAG:$Ce^{3+}$ target. They are studied by transmission electron microscopy (TEM), X-ray photoelectron spectroscopy (XPS), cathodoluminescence, near-field scanning optical microscopy (NSOM) and fluorescence in the photon-counting mode. Particles of extreme photo-stability as small as 10 nm in size are observed. These emitters are validated as building blocks of active NSOM tips by coating a standard optical tip with a 10 nm thick layer of YAG:$Ce^{3+}$ particles directly in the LECBD reactor and by subsequently performing NSOM imaging of test surfaces.

## 1. Introduction

Fluorescent NPs are currently one of the most interesting fields of research in materials science and modern optics. In particular, rare-earth doped nanocrystals which exhibit interesting optical properties in the visible and near-infrared [1, 2] are the focus of an increasing number of studies. Their high emission quantum yield [3], small size, photo-stability, and in the case of YAG:$Ce^{3+}$ (YAG matrix doped with cerium atoms in substitution to yttrium ions; YAG= $Y_3Al_5O_{12}$), short transition lifetime due to its fast parity-allowed transition d-f (< 70 ns in bulk [4] and ~ 30 ns for a NP [5]), have made rare-earth doped nanocrystals suitable candidates for a large variety of applications. For example, thanks to their high stability under irradiation, they are used in scintillators or displays [6], whereas YAG:$Ce^{3+}$ with its quantum yield in excess of 75% is used in white luminescence conversion light-emitting diodes [3]. Oxide NPs being in addition chemically inert, they can be used in biological applications [7, 8].

In parallel, the challenge of probing the electromagnetic field at the nanoscale thanks to NSOM or related nano-optical techniques has stimulated attempts of developing new optical nanosources or active tips. Active tips consist in the combination of a classical optical probe and a fluorescent nano-object, fixed at the apex, which plays the role of a light source. In principle, the lateral resolution should scale down to the size of the emitting particle, and should permit to perform, *e.g.*, fluorescence-resonance energy transfers controlled at the nanoscale [9, 10], high-resolution probing of the electromagnetic field lying in the vicinity of NPs or surfaces [11] and very accurate addressing of nano-objets or plasmonic devices [12].

Different emitters have been used to realize active tips such as a single terrylene molecule in a p-terphenyl micro-crystal [13], colour centres in thin LiF layers [14] or in diamond crystals [15], rare-earth-doped glass microparticles [16] and CdSe quantum dots (QDs) [17, 18, 19]. Work with a single molecule has been very successful, but it implies the use of cryogenic temperatures to prevent photo-bleaching. Nanodiamonds and LiF layers work at room temperature but they have offered a limited resolution so far because the emitter-to-sample distance (or equivalently the size of the hosting material) or the number of emitters involved in the optical signal are difficult to optimize. In addition, in the case of semiconductor QDs [20], either blinking and photo-bleaching or difficulties to fix one particle at the apex have limited the range of applications.

Therefore, there is need for alternative active objects that could be used as source of light in optical tips. Recently, colour-centre (NV) doped nanodiamonds with sizes around 25 nm have been pointed out as a promising alternative [21] due to their spectacular size reduction as compared to previous works [15]. However, their size is still slightly too large and their controlled grafting at the optical tip apex remains an open issue [22]. The present paper deals with another alternative which, apart from a recent report on the functionalization of atomic-force-microscope (AFM) tips with fluorescent $Gd_2O_3$:$Eu^{3+}$ nanorods [23], has not been reported so far in the context of optical tips, namely rare-earth doped oxide nanoclusters. We describe the synthesis of fluorescent oxide clusters by LECBD using a YAG:$Ce^{3+}$ target with a doping level of 1% of cerium ions substituted to yttrium ions. These NPs are studied by different methods including surface morphology by TEM and XPS and optical characterisation by cathodoluminescence, fluorescence photon-counting and NSOM imaging. With NSOM, we detect fluorescent NPs as small as 10 nm which demonstrates the possibility of working with small volumes of emitters. These NPs are remarkably photo-stable. We further present high-contrast NSOM images of test samples using, as source of light, the fluorescence light generated by YAG:$Ce^{3+}$ NPs deposited by LECBD directly onto the aperture of a metal-coated fibre tip. This validates the use of these oxide NPs as fluorescent medium in active optical tips.

2. **Experimental methods**

Rare-earth doped oxide NPs are synthesized in the gas phase with the LECBD technique [24]. A YAG target doped with 1% $Ce^{3+}$ ions in substitution to Y atoms is formed by pressing stoichiometric YAG:$Ce^{3+}$ powder and heating in a furnace up to 1293 K at air for 8 hours. This target is ablated by a pulsed (10 Hz repetition rate) Nd-YAG laser operating at 532 nm. The resulting plasma is first cooled by a continuous flow of neutral gas (25 mbar of helium) which initiates the nucleation of NPs. The synthesis is completed when the plasma undergoes a supersonic expansion (adiabatic expansion with a cooling rate of ~$10^{10}$ K/s) by flowing from the high pressure region where it is generated to the low pressure chamber (a few $10^{-6}$ mbar) where the clusters are deposited. The size of the resulting clusters produced by LECBD usually ranges between 5 and 10 nm [24].

Structural characterisations, such as crystallinity, cluster size and chemical composition, are carried out using TEM imaging and XPS measurements. TEM acquisitions are performed with a TOPCON microscope particularly adapted for high resolution imaging. XPS measurements are performed on a cluster assembled film with a CLAM 4 VG apparatus with a mean resolution of 0.2 eV, giving us information on the stoichiometry of the clusters. XPS is performed *in situ* after LECBD deposition.

Cathodoluminescence characterisation of clusters is performed *in situ* using a 5 keV electron probe with a spot size of approximately 1 mm$^2$. The electron current is set at 0.5 µA. The emitted light is collected by an optical fibre and sent to a Jobin Yvon TRIAX 320 monochromator preceding a cooled CCD detector with a 0.2 nm resolution.

Fluorescence is carried out on a home-made microscope that can operate both in confocal or NSOM transmission modes under ambient conditions [20]. The 4f-5d transition of cerium is excited with the 458 nm line of an Ar-Kr cw laser. The fluorescence light is collected from the sample through a ×60, NA 0.95 dry microscope objective and injected into a multimode optical fibre (core diameter: 50 µm) that is connected to an avalanche photodiode detector (SPCM-AQR 16, Perkin-Elmer, Canada). To remove the remaining excitation light, we use a dichroic mirror at 458 nm and a long pass filter transmitting light above 488 nm. Images are acquired with this microscope by raster scanning the sample under the optical tip (NSOM), whose particular characteristics are dictated by the envisioned measurements (see below), or the diffraction-limited focused laser beam (confocal).

## 3. Experimental results and discussion

### 3.1 Structural study

The XPS analysis of a few layers of oxide NPs on silicon gives us information of the exact stoichiometry of the sample. We observe peaks corresponding to the $2p_{1/2}$ and $2p_{3/2}$ levels of Al, $3d_{3/2}$ and $3d_{5/2}$ levels of Y and 1s level of O, respectively. However, we do not identify any signal corresponding to the trivalent state of cerium. We assume that the contribution of cerium is too small to be detected because of the small size of the clusters and the low doping level. These data confirm the presence of the elementary constituents of YAG in the NPs. The synthesized compound has a composition very close to the target, though slightly richer in oxygen. This small deviation in the stoichiometry appears during the critical step of nucleation in the LECBD reactor. This can mainly be explained by the involvement of four chemical elements, which renders the nucleation and the synthesis more complex than, *e.g.*, for $Gd_2O_3$ clusters [25] where there are fewer constituents.

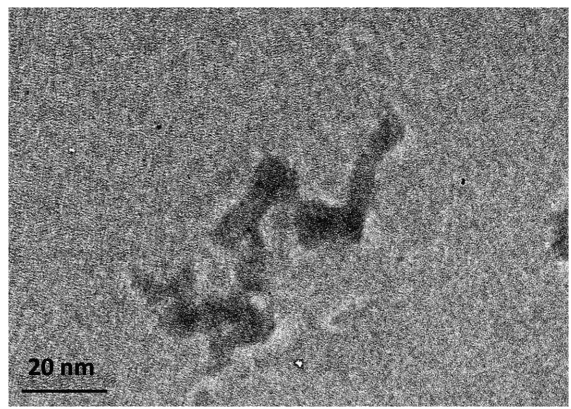

**Figure 1**. TEM image of aggregated YAG:$Ce^{3+}$ clusters synthesized by LECBD.

The TEM image of oxide NPs of figure 1 illustrates that their size ranges from 5 to 10 nm (consider the width of the wires) in agreement with other systems produced by LECBD, *e.g.*, $Gd_2O_3$ clusters [26]. Due to their ionic nature and, hence, to the dipolar interaction between each other during synthesis, the clusters are often aggregated on the surface in small numbers and take the shape of bent wires. We do not observe crystallized NPs. This is a direct consequence of the slightly difference of stoichiometry between the target and the clusters.

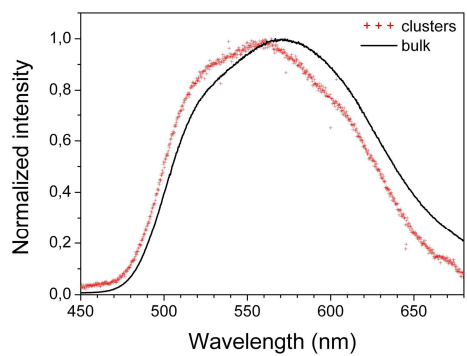

**Figure 2.** Normalized cathodoluminescence spectra of a few clusters monolayer (red crosses) and of the initial YAG:$Ce^{3+}$ target (black curve).

### 3.2 Optical studies

The luminescence properties of cerium in these oxide NPs are characterized and compared with the cerium emission in the YAG target. It is well known that the $Ce^{3+}$ luminescence in incomplete garnet surroundings is quenched due to the different behaviour of the lattice [27]. Though their composition varies slightly from the target, the oxide NPs synthesized here have a very similar emission spectrum as shown in figure 2 (the emission intensities are discussed further in the paper). This can be explained by the presence of crystallographic sites identical, or very similar, to those in the initial YAG target, which give rise to the $Ce^{3+}$ emission.

emit light thanks to a 5d-4f transition, where the lifetime of the excited level is much shorter (70 ns in the bulk *versus* ~ 1 ms for terbium for example). This characteristic makes cerium an excellent candidate as emitter in the YAG matrix. However, unlike 4f energy levels which are well shielded by filled 5s and 5p electronic shells, the 5d energy level is tremendously sensitive to the crystal field. For example, to illustrate the importance of the environment, cerium emits in the UV when it is embedded in a YAP ($YAlO_3$) matrix [28] and is a red emitter in sulfides [29].

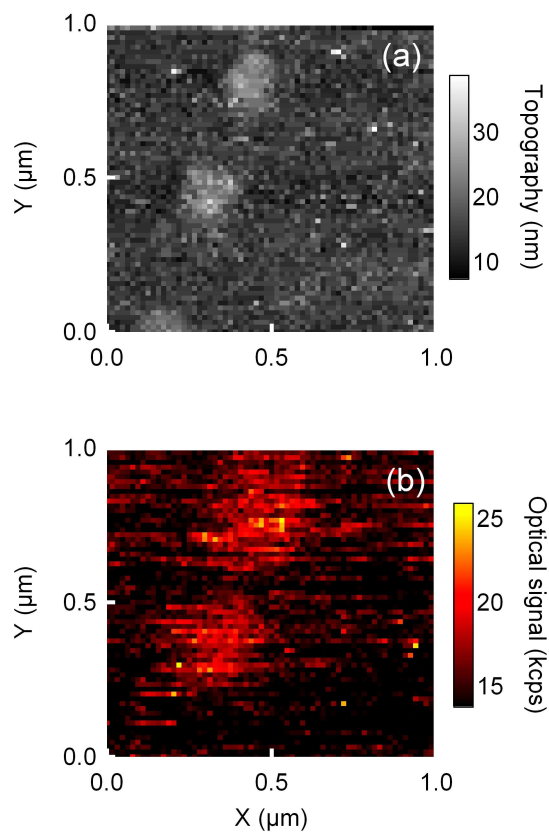

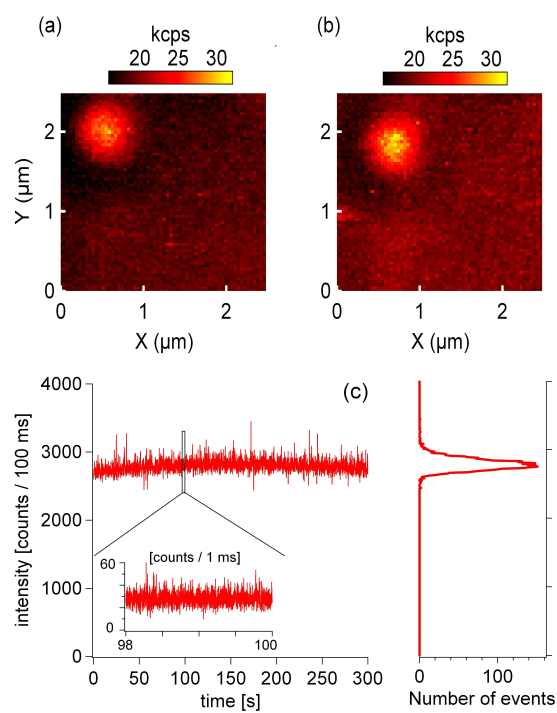

**Figure 3.** (a) Topography image (numerically flattened) and (b) near-field optical image of oxide NPs deposited on a fused silica cover slip (optical power at the uncoated tip apex: 23µW, integration time: 150 ms, scan speed: 0.5 µm s$^{-1}$, image size: 64x64 pixel$^2$). Both images are acquired simultaneously. The lateral (vertical) resolution in (a) is estimated to be 80 nm (2 nm). The spatial resolution in (b) is about 250 nm. The NPs are 10 nm across as estimated from their height in (a).

Figure 2 reveals a blue shift of the cluster emission which is centred at ~ 555 nm. This phenomenon is due to the electronic configuration of the cerium trivalent state. Unlike other rare-earth ions, where fluorescence originates from inter-4f forced transitions, cerium ions in the trivalent state

**Figure 4.** (a) and (b) Two successive fluorescence confocal images of a single YAG:$Ce^{3+}$ NP taken at a 30 minutes time interval (image size: 64 x 64 pixel², integration time: 70 ms, excitation power: 2 mW). The apparent different locations of the NP are due to thermal drifts. (c) Time trace of the NP emission taken after recording the image in (b) (binning time: 100ms, same power as above). The inset shows a zoom over 2 s (binning time = 1 ms). The plot on the right shows a histogram of fluorescence counts collected from the NP over the entire 300 s period.

Additional information on the optical properties of the new NPs is obtained from NSOM which has the ability of combining topographical and optical information at high spatial resolution. Figure 3 shows NSOM images of NPs sprayed by LECBD on a fused silica cover slip. Here, we used an uncoated

(not metallised) tip chemically etched from a single-mode optical fibre [30]. The use of an uncoated tip allows achieving a higher topographical resolution compared with a blunt metal-coated tip, though at the price of a reduced optical resolution. On figure 3, the sample has been heated at 500°C at air for 1 hour in order to aggregate the nanoclusters and turn the wire-like structures of figure 1 into more spherical (isotropic) ones, while leaving some of the NPs well isolated. This is favourable to study single NPs. As a consequence of the heat treatment, large structures up to more than 100 nm in height are indeed observed. However, remaining individual NPs as small as 10 nm across can still be detected. Interestingly enough, all small NPs detected in topography are also detected in fluorescence.

A useful optical signal of a few thousands counts per second (kcps) originating from the NPs is recorded (~ 4-5 kcps) under rather standard experimental conditions. In comparison, core-shell CdSe-ZnS nanocrytals emit a signal that can be ten times more intense under similar experimental conditions, but their photo-stability is too limited for the purpose of active-tip applications [20]. Colour centres in nanodiamonds also present very favourable properties for the latter purpose and are even brighter than the oxide NPs but their size around 25 nm reported so far remains slightly larger [21].

A stringent test of the YAG:$Ce^{3+}$ NPs concerns their photo-stability. To address this issue, we use the confocal mode of our fluorescence microscope which is less sensitive to thermal drifts than NSOM, thereby allowing for long-term measurements on a fixed single NP. Figures 4(a) and 4(b) depict two fluorescence images of the same single NP taken at a 30 minutes time interval. During this period of time, the NP has been continuously illuminated for 10 minutes with a focused laser power of 2 mW, corresponding to a huge excitation density in excess of 100kW/$cm^{-2}$. As can be seen, the NP remains insensitive to this rather severe treatment in contrast with, *e.g.*, CdSe QDs which rapidly photo-bleach at lower excitation density [31]. As a further proof for the absence of photo-bleaching, we note that we could build repeated NSOM or confocal fluorescence images of regions of samples stained with single YAG:$Ce^{3+}$ NPs without observing any one ever bleaching.

The former experiments give us an opportunity to check the photo-stability of our oxide NPs against fluorescence blinking which is another hallmark of the photo-physics of single molecules and semiconductor QDs [17, 32]. In figure 4(c), the PL intensity of the single NP seen in figure 4(b) is plotted as a function of time. Here, scanning has been stopped, in such a way that the laser spot is focused on the NP. It can clearly be seen that under continuous illumination the studied NP does not blink whatsoever, neither over the whole examination period of 300 s, nor over the zoomed period of 2 seconds (the slow change in the signal level is due to thermal drifts of ~ 5 nm/minute). This non-blinking nature of the NP emission is confirmed by the histogram of fluorescence counts given on the right hand side of figure 4(c), which can be fitted with a single Gaussian curve (half-width at half-maximum: 226 count/100 ms). This is in contrast with single CdSe QDs which exhibit two well-defined peaks corresponding to the on- and off-periods of their blinking emission (see e.g. Ref [31]). This remarkable photo-stability of the YAG:$Ce^{3+}$ NPs agrees with the facts that a large number (above 100) of deep emitters are involved in the fluorescence signal of a 10 nm 1%-cerium-doped YAG NP and that YAG, being an oxide by nature, should not be sensitive to photo-oxidation processes.

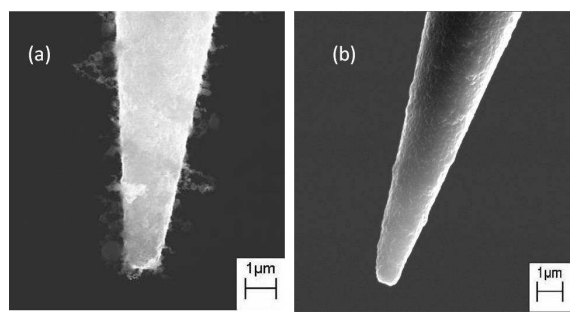

**Figure 5**. (a) Electron micrograph of a metal-coated optical tip subsequently loaded with YAG clusters deposited directly in the LECBD chamber. (b) An electron micrograph of a bare metal-coated tip shown for comparison.

*3.3 Active tips*

The possibility of working with very small emitting volumes of a stable temporal behaviour leads us to consider using the oxide NPs as light source for active tips. To take advantage of the high vacuum inside the LECBD chamber, we introduce here a method for realizing active tips which has not been considered previously to the best of our knowledge.

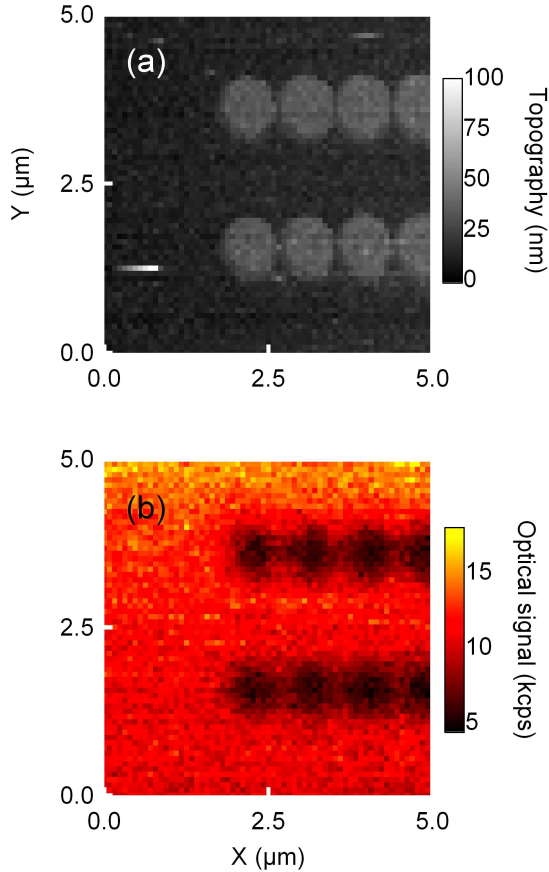

**Figure 6.** (a) Numerically flattened topography image and (b) near-field optical image acquired simultaneously on a sample consisting in gold discs (400 nm diameter) deposited on a fused silica cover slip. The probe is a 400-nm metal-coated tip functionalized with oxide NPs. Optical power at the tip apex: 1.4μW, integration time: 20 ms scan speed: 1 μms$^{-1}$.

We start from an aluminium-coated optical tip [33] which potentially offers a tighter light confinement than uncoated tips. This substrate tip has been first characterized for lateral light leakage and effective aperture size, found to be 400 nm, by measuring its far-field angular intensity distribution [34]. This tip is introduced into the LECBD chamber.

Then, YAG:Ce$^{3+}$ NPs are sprayed directly onto the tip apex by LECBD. The LECBD reactor parameters are set so as to grow a layer of an equivalent thickness of 10 nm. Figure 5 shows a Scanning Electron Microscope (SEM) image of such a coated tip. Obviously, the clusters do not cover homogenously the tip surface but they aggregate locally under the shape of thin wires, the first clusters deposited on the rough aluminium surface playing the role of nucleation centres.

To validate YAG:Ce$^{3+}$ coated tips as active optical tips, NSOM imaging is performed in the transmission mode by using the light emitted in the spectral range (see figure 2) of the YAG:Ce$^{3+}$ clusters as source of light (no heat treatment has been applied to the active probe). A test sample with opaque gold discs of 400 nm in diameter deposited on a glass cover slip is used. Optical filtering allows to restrict the light collection to the spectral range of YAG:Ce$^{3+}$. The images shown on figure 6 confirm that the light emitted by the functionalized tip is intense enough to reveal the gold pattern as non-transmitting dark spots with a good contrast. The lateral resolution inferred from the optical image is of the order of the substrate aperture size only, which indicates that the entire free aperture is covered with oxide clusters (note that NPs deposited on the metallic sides of the tip do not contribute to the emission signal). The useful signal level of almost 6 kcps collected through the transparent parts of the test sample is a sizeable one. Because it is obtained under rather standard experimental conditions, this is encouraging to envision improved resolutions by using oxide layers of smaller lateral extent. Routes towards this goal may combine the use of a larger cerium doping level, mass-selected cluster deposition [35], appropriate shaping of the substrate tip by focused-ion beam milling [36], and higher excitation rates thanks to the exceptional photo-stability of oxide NPs.

## 4. Conclusion

Fluorescent YAG:Ce$^{3+}$ oxide NPs as small as 10 nm in size have been synthesized following the LECBD physical route. These particles are remarkably photo-stable at room temperature and exhibit no blinking and no bleaching whatsoever. Preliminary NSOM imaging with a tip coated with these oxide NPs directly in the LECBD chamber confirms the potential of the new NPs of acting as nanosources of light in NSOM with active optical tips.

**Acknowledgments**

We are grateful to Jean-François Motte for the SEM imaging and the preparation of optical tips, and to Olivier Boisron for fruitful discussions about XPS measurements. The PhD grant of A Cuche and partial support to this research by the Région Rhône-Alpes ("Cluster MicroNano") are gratefully acknowledged.